\documentclass[11pt]{article}
\begin{document}

\def \ov{\over}
\def \bar{\overline}
\def \beq{\begin{equation} }
\def \eeq{\end{equation} }
\def \lb{\label}
\def \nn{\nonumber}
\newtheorem{lemma}{Lemma}
\newtheorem{theorem}{Theorem}
\newtheorem{proposition}{Proposition}
\newtheorem{definition}{Definition}

\baselineskip .54cm
\def\nn{\nonumber}

\def \pd{\partial}
\def\~#1{\widetilde #1}
\def\.#1{\dot #1}
\def\^#1{\widehat #1}
\def\d{{\rm d}}       
\def \id{\! :=}
\def\dst{\displaystyle}

\def \sy {symmetry}
\def \sys {symmetries}
\def \so {solution}
\def \eq{equation}
\def \R{{\bf R}}
\def \vf{vector field}
\def\tr{transformation}

\def\a{\alpha}
\def\be{\beta}
\def\phi{\varphi}
\def\de{\delta}
\def\g{\gamma}
\def\De{\Delta}
\def\th{\theta}
\def\ka{\kappa}
\def\s{\sigma}
\def\la{\lambda}

\def \qq{\qquad}
\def \q{\quad}
\def \pn{\paragraph\noindent}
\def \sk{\medskip}
\def \ni{\noindent}
\def\CS{conditional symmetry}
\def\eu{({\ref{EU})}}
\def\={\, =\, }

\title{{\bf Symmetry properties of some Euler-type equations in plasma
physics}}

\author{F. Ceccherini\thanks{Email: ceccherini@df.unipi.it} $^{(a,b)}$,
   G. Cicogna\thanks{Email: cicogna@df.unipi.it} $^{(a,c)}$
   and F. Pegoraro\thanks{Email: pegoraro@df.unipi.it} $^{(a,b)}$ \\
   \\~\\
$^{(a)}$ Dipartimento di Fisica ``E.Fermi'' dell'Universit\`a di Pisa\\
   $^{(b)}$ Istituto Nazionale di Fisica della Materia, Sez. A\\
   $^{(c)}$ Istituto Nazionale di Fisica Nucleare, Sez. di Pisa \\~\\
Largo B. Pontecorvo 3, Ed. B-C, I-56127, Pisa, Italy  }


\maketitle

\begin{abstract}
We consider a system of partial differential equations, of interest
to  plasma physics, and provide all its Lie point symmetries,
with their respective invariant
solutions. We also discuss some of its conditional and partial
symmetries. We finally show that, although the system can be
cast in divergence form and admits conserved currents, it does not 
admit potential symmetries.
\end{abstract}


\vfill\eject

\section*{Introduction}
It is well known that the analysis of \sy\
properties is one of the most interesting tools in the study of
both ordinary and partial differential \eq s,
and of systems thereof (see \cite{Ov}-\cite{BA} and references therein);
it is also known that this  analysis can be of concrete help in finding
explicit \so s to these \eq  s. It is impossible to provide a fairly complete list 
of references devoted to these ideas; let us only mention here a very recent 
paper \cite{Bi}, where \sy\ methods are used in a problem of applied 
mathematics.

At present, finding \sys\ of a given differential \eq\ is an
(almost)
completely standard routine, thanks also to some computer packages (see
e.g. \cite{Her,Ba}) which can help in the calculations (we will consider
only Lie point-\sys , see below). While this is true for what concerns
{\em exact} \sys , this may be not completely obvious when the more
refined and equally useful notions of {\em conditional} or {\em partial}
\sys\ are introduced (their definitions will be recalled  in a moment).

Although these notions are not new,  the system of  partial differential
\eq s which we are  going to analyze, and which is of  interest to
plasma physics and   has been widely used recently in the investigation
of ``magnetic field line reconnection''(see \cite{P1}-\cite{P4}), is so rich
of various \sy\ properties and admits so  different  types of nontrivial
\sys\ and particular \so s, that  we  believe that the application of \sy\ methods 
to this problem deserve to be  proposed both to specialists in plasma physics 
and to experts in \sy\  theory. For a different application of \sy\ methods to
the analysis of some \eq s (the Vlasov \eq s) relevant in plasma physics, we refer
to \cite{K1,K2}.

Let us briefly recall (see e.g. \cite{Ol} for details) that,
given a system $\De\equiv \De_\ell (x,u^{(m)})=0$ of $\ell$ partial
differential \eq s for the
$q$ dependent variables (or unknown functions) $u\equiv u_a(x)$
($a=1,\ldots,q$) of the $p$ independent variables $x\equiv
(x_1,\ldots,x_p)$ ($u^{(m)}$ denotes the derivatives of $u$ up to some
order $m$), a \vf\ $X$
\beq X\= \xi_i(x,u){\pd\ov{\pd
x_i}}+\zeta_a(x,u){\pd\ov{\pd u_a}}\lb{XG}\eeq
(sum over repeated indices)
is the Lie generator of an (exact) \sy\ for the system $\De=0$ if the
appropriate prolongation $X^*$ of $X$ satisfies
\beq X^*(\De)|_{\De=0}\= 0
\lb{XD} \eeq
i.e. if $X^*(\De)$ vanishes along the \so s of $\De=0$. If
$X$ is a  \sy , then -- once any \so\ to $\De=0$ is given -- one may
obtain, under the action of $X$, an orbit of \so s to $\De=0$. There may
exist also special \so s $u^0=u^0(x)$ which are left fixed under $X$:
these satisfy the invariance condition
\beq X_Qu^0\= 0\lb{XQ} \eeq
where
\beq X_Q=\Big(\zeta_a-\xi_i{\pd u_a\ov{\pd x_i}}\Big){\pd\ov{\pd u_a}}
\lb{XQa}\eeq
is the \vf\ $X$ written in ``evolutionary form'' \cite{Ol}.

In addition to these (exact) \sys , we will also consider weaker notions
of \sys , namely the conditional \sys\ and the partial \sys . Roughly
speaking, conditional \sys\ are given by \vf s which admit only invariant
\so s, as in (\ref{XQ}), whereas partial \sys\ are \vf s which transform
into other \so s of $\De=0$ only the \so s which belong to a proper subset
of all \so s (i.e. only the \so s satisfying some additional condition).
These notions will be more precisely defined in sect. 2 and 3
respectively.

\section{Exact \sys }

The system of partial differential \eq s we are going to
consider is the following

\smallskip

\beq\cases{
{\pd\over{\pd t}}(\psi-\De \psi +
\De\phi)+ \big[\phi+
\psi,\psi-\De \psi +
\De\phi \big]\ = 0    \cr
{\pd\over{\pd t}}(\psi-\De \psi -
\De\phi)+\big[\phi-
\psi,\psi-\De \psi
- \De\phi\big]\ =0    \lb{EU}}\eeq
here $x\equiv(x,y,t)$, $u\equiv (\psi,\phi)$, 
  $~\psi=\psi(x,y,t)$, $~\phi=\phi(x,y,t)~$ and
\[
[f,g]\= {\pd f\over{\pd x}}{\pd g\over{\pd y}}-{\pd g\over{\pd x}}{\pd
f\over{\pd y}}
\ .\]

The above equations describe the low-frequency  nonlinear evolution 
of a two-dimensional
plasma configuration embedded in a strong magnetic field in the
$z$-direction  and with a shear magnetic field in the
$x$-$y$ plane $~{\bf B}=B_0{\bf e}_z+{\bf \nabla}\psi\times{\bf e}_z.~$
These equations can be obtained from a generalized two-fluid 
dissipationless  model of the
plasma response  where the plasma velocity in the $x$-$y$ plane is given by
the electric drift $~{\bf v} = {\bf e}_z\times \nabla \phi$, $~\phi$ 
is proportional to the
electric  potential in the plasma  and plays the role of a stream 
function, while $~\psi$ is a
magnetic flux function proportional  to the $z$ component of the 
magnetic vector potential
\cite{P1}-\cite{P4}.   This model  includes the effects of the electron inertia
and of the electron pressure.    For the sake of simplicity here we have
adopted a suitable rescaling of the variables and set all physical
dimensionless parameters equal to one; see also Sect. 3.2.

There are some obvious
\sys\ of the above system, namely  spatial and time translations, and
spatial rotations, generated respectively by
$${\pd\ov{\pd x}},\q {\pd\ov{\pd y}},\q {\pd\ov{\pd t}},\q
y{\pd\ov{\pd x}}-x{\pd\ov{\pd y}}$$
Other trivial \sys\ are given by the \vf s
$${\pd\ov{\pd \psi}},\q q(t){\pd\ov{\pd \phi}}$$
which generate the
\tr s $\psi\to \psi+k,\ \phi\to\phi+q(t)$, where $k$ is a constant and
$q(t)$ an arbitrary function, which do not change either the
magnetic field or the plasma velocity.

With the help of some appropriate computer package, e.g.  \cite{Her,Ba},
it is possible to show the following result:

{\proposition $\!\!\!$.
Apart from the above mentioned, trivial, \sys , the system \eu\ admits
only the following \sys : the infinite dimensional subalgebra
\beq
X_1= A(t)  {\pd\over{\pd x}}+ B(t)
{\pd\over{\pd y}}+\Big(x {\d B\over\d t}  - y {d A\over{\d t}}
\Big){\pd\over{\pd\phi}}\lb{X1} \eeq
where $A(t),B(t)$ are (nonconstant) arbitrary differentiable functions,
and the vector field
\beq \lb{Xr}
X_2\= -t\, y\ {\partial\over {\partial x}}+t\, x\ {\partial\over{\partial
y}}+ {x^2+y^2\over {2}}\ {\partial \over{\partial \phi}} \lb{X2}\eeq
}

\medskip\ni
We will now consider in some detail these two cases.

\medskip\ni
1)  The subalgebra $X_1$ describes an infinite family of \sys , due to the 
presence of the functions $A(t),B(t)$. This expresses the property 
that, if $\psi(x,y,t),\ \phi(x,y,t)$ is a \so\
of \eu , then also\footnote{A continuous Lie parameter $\la$  should be
introduced to parametrize this family of \so s, but it can be clearly
absorbed in the (arbitrary) functions $A$ and $B$.}
\[ \Psi(x,y,t):=\psi(x-A(t),y-B(t),t) \  , \]
\[ \Phi(x,y,t):=xB_t-yA_t-{1\over
2}(AB_t-A_tB)+\phi\Big(x-A(t),y-B(t),t\Big) \]
($A_t=\d A/\d t$, etc.)
solve our system \eu , for any $A(t),B(t)$. It must be noticed that this
corresponds to a change of spatial coordinates into a moving frame which
produces in turns the additional term $xB_t-yA_t-(1/2)(AB_t-A_tB)$ in the component
$\phi$.   Since $\phi$ is proportional to the electric potential,
we can interpret this \sy\ as expressing
fact that  a time dependent, spatially uniform, electric field
imposed on the system induces a uniform time-dependent
electric drift  $A_t {\bf e}_x + B_t {\bf e}_y$.

We now look for the \so s to \eu\ which are {\it invariant}
under (\ref{X1}), i.e. for \so s to (\ref{EU}) satisfying the invariance
condition (\ref{XQ}), which now takes the form ($\psi_x=\pd\psi/\pd x$,
etc.)
\[ A\psi_x+B\psi_y=0\q  , \q A\phi_x+B\phi_y=xB_t-yA_t \ .\]
The first equation  can be interpreted as the requirement  that the
displacement $A {\bf e}_x + B{\bf e}_y$ produced by the electric drift
$A_t {\bf e}_x + B_t {\bf e}_y$ is parallel to the  field lines
of the shear magnetic field ${\bf \nabla}\psi\times{\bf e}_z$, while
the second  expresses the
requirement   the electric potential of the displaced plasma element
remains  constant.

The above \eq s may be easily integrated to get
\beq \lb{VW}
\psi=V(s,t) \, ,\,
\phi={1/2\over{A^2+B^2}}\Big((A_tB+AB_t)(x^2-y^2)-2xy(AA_t-BB_t)\Big)\,
+W(s,t) \eeq
where $s=B(t)x-A(t)y$ and $t$ are  $X_1$-invariant
variables, so that
$\nabla s$ is orthogonal to the plasma  displacement;
the configuration (\ref{VW}) corresponds to a magnetic field
aligned along the direction of the  displacement and to a time dependent
velocity pattern consisting of  a
hyperbolic field with stagnation point at $x= y = 0$ superimposed to
a   field  uniform along the  plasma displacement.

Substituting (\ref{VW}) into the system \eu\  we obtain
the following  ``reduced'' system for the functions
$V=V(s,t),W=W(s,t)$
\[ {\pd\over{\pd t}}\Big(V-(A^2+B^2)V_{ss}\Big)=0 \q , \q
{\pd\over {\pd t}}\Big((A^2+B^2)W_{ss}\Big)\= 0
\]
which clearly imply
\[ V-(A^2+B^2)V_{ss}\= F(s) \q ,\q
(A^2+B^2)W_{ss}\= G(s) \]
where $F(s),\, G(s)$  are arbitrary functions. 
Notice   that $\pd_{xx}+\pd_{yy}=(A^2+B^2)\,\pd_{ss}$
and that these \eq s are actually ODE's, indeed the variable $t$ here 
appears  merely as a parameter.

Special simple solutions with a uniform magnetic field and a purely 
hyperbolic velocity field can  be obtained with $V(s)=s$,
i.e. $\psi=B(t)x-A(t)y$, and $W=0$ in  (\ref{VW}).
In the elementary case $A(t) = \pm B(t) = A_0\exp{(t)}$   this solution 
corresponds to the
exponential  growth of the magnetic field amplitude in a time 
independent velocity field,
while in the case $A(t) = A_0\cos{(\omega t)}, B(t) = A_0\sin{(\omega 
t)}$ it corresponds to a
magnetic field rotating with frequency $\omega$ in a velocity field 
rotating with frequency $2 \omega$.

\smallskip

It is remarkable  that this reduced system
consists of two {\it linear and uncoupled} homogeneous \eq s; in particular,
it produces \so s where $\psi$ satisfies a {\it linear}
\eq\  (therefore linear superposition principle holds), and  $\phi$
does the  same, apart from a fixed additional term.
The fact that the invariant \so s describe a linear manifold may
appear rather surprising: we shall comment on this point at the
end of this section.

\bigskip\ni
2)
The other \sy\ $X_2$ [eq. (\ref{X2})]
does not depend on arbitrary functions; it implies
that  if $\psi(x,y,t),\ \phi(x,y,t)$ is a
\so\ of \eu , then also
\[ \lb{ppr}\Psi(x,y,t):=\psi\Big(x \cos(\lambda
t)+y \sin(\lambda t),\
-x\sin(\lambda t)+y \cos(\lambda t),t\Big) \]
   \[\Phi(x,y,t):=\phi\Big(x \cos(\lambda t)+y \sin(\lambda t),\
-x\sin(\lambda t)+y
\cos(\lambda t),t\Big)+\lambda{x^2+y^2\over{2}}\]
are a family of \so s to \eu\ for
any $\la\in\R$. This represents a sort of rotated \so s with angular
velocity $\la$ plus a radial term in the component $\phi$
 which gives the additional velocity field corresponding to the 
rotation. As a trivial
example, starting from the simple \so\ $\psi=\exp(-x^2) , \phi=1/(1+x^2)$
of \eu , we  can conclude that
\[\Psi(x,y,t):=\exp\Big(-(x\cos(\la t)+y\sin(\la t))^2\Big) \]
\[ \Phi(x,y,t):=\Big(1+(x \cos(\lambda t)+y \sin(\lambda
t))^2\Big)^{-1}+\lambda{x^2+y^2\over{2}} \]
also solve (\ref{EU}).

We now look for \so s to \eu\ which are {\it invariant} under  (\ref{X2}),
i.e.  for \so s to \eu\ satisfying the invariance condition (\ref{XQ}),
which now reads
\[ y{\pd
\psi\over{\pd x}}-x{\pd \psi\over{\pd y}}\= 0 \qquad , \qq
y{\pd
\phi\over{\pd x}}-x{\pd \phi\over{\pd y}}\=-{r^2\over {2t}} \]
where
$r^2=x^2+y^2$. It is easy to find that these $X_2-$invariant \so s to \eu\
are of the form, with $ \theta=\arccos(x/r) $,
\[ \psi=Q(r,t) \quad ,\quad
\phi={r^2 \over{2t}}\ \theta\ +R(r,t) \]
As in the above case,  substituting in \eu , one obtains that
the functions $Q$ and $R$ must satisfy  two {\em
linear} and uncoupled homogeneous \eq s:
$$ \begin{array}{l}
r^2Q_{rrr}-2rtQ_{rrt}+rQ_{rr}-2tQ_{rt}-r^2Q_r+2rtQ_t+3Q_r=0  \cr
   2rtR_{rrt}-r^2R_{rrr}+2tR_{rt}-rR_{rr}+5R_r=0
\end{array}$$
It can be noticed in particular the special form of the component $\phi$,
which has a fixed term with a ``cut'' discontinuity, which looks like a
 spiral and expresses the fact
that this solution contains a $\theta$-independent azimuthal electric field,
which  vanishes for $t\to\pm\infty$; whereas the \eq\ for the other term
$R(r,t)$ admits \so s of the form
$$R=t^a r^b \q\q {\rm with}\quad a={b^2-2b-4\over{2b}},\quad \forall b \ .$$

We can now show why all the \sys\ considered above yield
invariant \so s which belong to a linear manifold. This follows from this
simple result.

{\lemma $\!\!\!$.
Write the \vf\ $X$ (\ref{XG}) in the form
\beq \lb{Xex} X\= \xi{\pd\ov{\pd x}}+\eta{\pd\ov{\pd
y}}+\tau{\pd\ov{\pd t}}+\zeta{\pd\ov{\pd\psi}}+\chi{\pd\ov{\pd \phi}} \eeq
Assuming that
(\ref{Xex}) is a projectable \vf\ (i.e. that $\xi,\eta,\tau$ do not depend
on $\psi,\phi$), and assuming also that $\zeta$ does not depend on
$\phi$, and $\chi$ on $\psi$, the general $X$-invariant \so s  to
\eu\ take the form
\beq \lb{ZZ} \psi=\a_1(x,y,t)+Z_1(s_1,s_2)\ , \
\phi=\a_2(x,y,t)+Z_2(s_1,s_2)\eeq
where $s_1,s_2$ are $X$-invariant coordinates, $\a_1,\a_2$ are fixed
functions,
and $Z_1,Z_2$ satisfy linear \eq s if $[s_i,s_j]=0$.}

\medskip\ni
{\it Proof}. The proof comes from a simple calculation.
The general invariant \so s under $X$ can be found solving the
characteristic \eq
\beq \lb{ch} {\d x\ov{\xi}}={\d y\ov{\eta}}={\d
t\ov{\tau}}={\d\psi\ov{\zeta}}=
{\d\phi\ov{\chi}}\eeq
Now, the invariant coordinates $s_1,s_2$ are obtained from the subsystem
${\d x/{\xi}}={\d y/{\eta}}={\d t/{\tau}}$, $\a_1,\a_2$ are  functions
determined by the remaining
\eq s in (\ref{ch}), and finally $Z_1,Z_2$   must satisfy the
reduced \eq s obtained substituting (\ref{ZZ}) into \eu . It is now
easy to see that the nonlinear terms for the \eq s containing $Z_i\
(i=1,2)$ come from $[Z_i,Z_j]$ and $[Z_i,\De Z_j]$ in \eu , and a direct
calculation shows that these quantities are proportional to $[s_i,s_j]$.
$\hfill\triangle$

\bigskip
Therefore, if e.g. one of the invariant variables $s_i$ can be chosen to
depend  only on the time $t$, as in the cases considered above, all
nonlinear terms in the reduced \eq s disappear and the invariant \so s
have the form, as given in (\ref{ZZ}), of a fixed function $\a_i$ plus a
term $Z_i$  belonging to a linear space.

\medskip
As seen in the above discussion, there are mainly
two applications of \sy\ properties of a differential problem, i.e.
finding new \so s starting from a know one, and looking for invariant \so
s under the \sy . The introduction of the conditional and partial \sys ,
which are {\it not} exact \sys\ of the problem, has essentially the scope
of preserving just one of these two features of exact \sys .

\section{Conditional \sys }

For ``historical'' reasons, we will
consider first the case of conditional \sys\ \cite{BC}-\cite{FK} ,
although in the
case of our problem \eu , more interesting situations occur in the
presence of partial \sys\ (which will be considered in next section).

A conditional \sy\ for an \eq\ (or system of \eq s) $\De=0$ is -- roughly
--
a \vf\ $X$ which is a \sy\ of the enlarged system obtained appending to
the initial \eq\ $\De=0$  the \eq\ expressing the invariance under
$X$, i.e. $X_Qu=0$, as in (\ref{XQ}). It is clear that if $X$ is a
conditional \sy\ (and not an exact \sy ), there are no (proper) orbits of
\so s, indeed conditional \sys\ do not map, in general \so s of $\De=0$
into other \so s, but are introduced in order to find just $X$-invariant
\so s.

There are some delicate points related to this definition: see
\cite{OR}-\cite{ZTP}.  In particular, it has been pointed out \cite{OR}
that, given an \eq\
$\De=0$, {\it any} \vf\ $X$ can be a conditional \sy , the only condition
to be verified is that the system $\De=0,\ X_Qu=0$ does admit some \so .
Conversely, one can say that {\it any} \so\ to $\De=0$ can be considered
an invariant \so\ under some \vf\ $X$. This remark is actually related to
the introduction of a subtler classification of the notion of conditional
\sys\ \cite{CiK}, which will not be considered here, and which depends on
the number of auxiliary \eq s (differential consequences) needed for
solving the problem. Therefore, if one is not interested in this
classification, but only in finding \so s of the problem $\De=0$, one has
just to choose ``reasonable'' \vf s $X$ and check if the system
$\De=X_Qu=0$ admits \so s; these \so s, by construction, are invariant
under $X$.

It often happens that the \so s found in this way are trivial
(e.g. $u=$ const), or may be obtained by means of different procedures.
E.g., considering for our system \eu\  the case of spatial
dilations (scalings)
$$X\=a\, x\, {\pd\ov{\pd x}}+b\, y\, {\pd\ov{\pd
y}}\q\q a,b={\rm const}$$
which  are not exact \sys , we get the
disappointing result that the only \so\ to \eu\ and $X_Qu=0$ must be
independent of the time, $\psi_t=\phi_t=0$, and must satisfy
$\psi=\pm\phi$. Unfortunately, this does not produce useful indications,
because {\it any} couple of functions $\psi_0(x,y)$ and
$\phi_0(x,y)=\pm\psi_0(x,y)$ are \so s to $\De=0$, although not scaling
invariant~!

A more interesting situation occurs considering the \vf
\beq
X\=\psi_y{\pd\ov{\pd \psi}}+\phi_x{\pd\ov{\pd\phi}}\lb{X3}\eeq
which is actually a \vf\ in evolutionary form not
reducible to a \vf\ in standard form (\ref{XG}) (or a ``contact \sy ''
\cite{Ol}). The $X$-invariant \so s  of \eu\ which can be found are the
following
$$\psi=\sin[k \,
(x-\gamma(t))] \q , \q
\phi=\exp[\pm(1+k^2)^{1/2} y]-{\d\gamma\over{\d t}}\, y+T(t)$$
or
$$\psi=\exp[ \pm(1+k^2)^{1/2}  \ (x-\gamma(t))] \q , \q
\phi=\sin(ky)-{\d\gamma\over{\d t}}\, y+T(t)$$
where $\gamma$ e $T$ are arbitrary functions and $k\in\R$,
or also, with $|\kappa|<1$,
$$\psi=\exp[\pm(1-\kappa^2)^{1/2}(x-\gamma(t)] \q , \q
\phi=\exp(\pm\kappa
y)-{\d\gamma\over{\d t}}\, y+T(t)$$
Notice in particular the presence of
terms depending on $x-\gamma(t)$ describing a generalized wave
propagation.
Other \so s which can be obtained in the same way are
$$\psi=\gamma(t)\exp(\pm x)\q , \q \phi=k\, y^2+y\, T_1(t)+T(t)$$
($\gamma,\ T_1$, $T$ arbitrary functions), and
$$\psi=\gamma(t)+kx\q , \q \phi=\exp(\pm y)+{\d\gamma\over{\d t}}{y\over
k} \qquad (k\not= 0)$$
and similar \so s of the same form, obtained
changing $sin$ into $cos$, adding constant terms and so on.

Needless to
say, starting from these \so s and using the exact \sys\ of \eu\ examined
in Sect. 1, one may construct other families of \so s  to \eu .

\section{Partial \sys . }

The notion of partial \sy\ has been introduced in \cite{CG} (see also
\cite{CC}). As already  mentioned, while exact \sys\ transform any \so\ of
the
given problem into  another \so , partial \sys\ do the same only for a
proper
subset of \so s,  which is defined by some supplementary differential \eq
s.
More precisely, let us now  assume  that $X$ is {\it not} an exact \sy ,
therefore $X^*(\De)|_{\De=0}\not= 0$;  let us then introduce the condition
\beq \De^{(1)}:=X^*(\De)\= 0\lb{D1}\eeq
as a new \eq , and consider the enlarged system
\beq
\De=\De^{(1)}\= 0\lb{DD} \ . \eeq
It is clear that if $X$ is an exact \sy\
of this system, then the subset ${{\cal S}}^{(1)}$ of the simultaneous \so
s of this system is a ``symmetric set of \so s'' to $\De=0$, i.e. a proper
subset of \so s which have the property of being mapped   the one into
another by the \vf\ $X$. It is also clear that this property is not shared
by the other  \so s to $\De=0$ not belonging to the subset ${{\cal
S}}^{(1)}$. In  principle, this procedure can be iterated introducing, if
$X$
is not a
\sy\ of the enlarged system (\ref{DD}), further \eq s
$\De^{(2)}:=X^*(\De^{(1)})=0$ and so on, but this possibility will not be
considered here.

\subsection{Example 1}
We will now introduce a first example of partial \sy\
for the problem \eu . Consider the \vf\ of the form
\beq X_{ab}= -t\, y\ {\partial\over {\partial x}}+t\, x\
{\partial\over{\partial
y}}+  {x^2+y^2\over {2}}\ \Big(a\ {\partial \over{\partial \psi}}+b\
   {\partial \over{\partial \phi}}\Big) \lb{ab}\eeq
where $a,b$ are
constants (not both zero), which is similar to (\ref{X2}) but now involves
also the component $\psi$. According to our procedure, let us evaluate
$X_{ab}^*(\De)$: we obtain a system of \eq s which can be rewritten in
this form
\beq \De^{(1)}:=X^*_{ab}\De=
\cases{
\Big(y{\pd\ov{\pd_x}}-x{\pd\ov{\pd y}}\Big)
\Big((1-b)(\psi-\De\psi)+a(\phi-\De\phi)\Big)=0\cr
\Big(y{\pd\ov{\pd_x}}-x{\pd\ov{\pd y}}\Big)
\Big(a\De\psi+(1-b)\De\phi\Big)=0}   \lb{D1e} \eeq
It is clear that $X_{ab}^*(\De)=0$ if and only if $a=0,b=1$, which
corresponds
to the exact \sy\ (\ref{X2}) already considered in Sect. 1. Excluding 
this case, it is not difficult to show that the enlarged system
(\ref{DD}), which is now given by the  the four \eq s \eu\ and (\ref{D1e}),
admits $X_{ab}$ as an exact \sy , i.e.
$X^*_{ab}(\De)=X^*_{ab}(\De^{(1)})=0$ when $\De=\De^{(1)}=0$. Therefore,
$X_{ab}$ is a partial \sy , and the set ${{\cal S}}^{(1)}$ of the
simultaneous
\so s of the  system \eu , (\ref{D1e}) (which is clearly a subset of all
the
\so s of $\De=0$, i.e. of \eu ), is symmetric under $X_{ab}$, and --
as said before
-- the \so s belonging to this set are transformed by $X_{ab}$ into other
\so s to \eu . Written explicitly, if $\psi(x,y,t),\phi(x,y,t)$ solve \eu\
and
(\ref{D1e}), then
\[ \Psi(x,y,t):=\psi\Big(x\cos(\la t)+y\sin(\la t),-x\sin(\la t)+y\cos(\la
t),t\Big)+a\la{x^2+y^2\ov{2}}\]
\[ \Phi(x,y,t):=\phi\Big(x\cos(\la t)+y\sin(\la t),-x\sin(\la t)+y\cos(\la
t),t\Big)+b\la{x^2+y^2\ov{2}}\]
also solve \eu , for all $\la\in\R$.

Different choices of the constants $a,b$ give different interesting
possibilities; for instance:

\smallskip\ni
i) if $a=0,b\not=0,1$, then the additional \eq\ $\De^{(1)}=0$ is satisfied
by functions $\psi,\phi$ such that
\beq \psi-\De\psi\= F_1(r,t)\q ,\q \De\phi\=F_2(r,t)
\lb{ci}\eeq
where (here and in the remainder of this subsection) $F_i(r,t)$
denote arbitrary functions of $r=\sqrt{x^2+y^2}$ and the time.

\smallskip\ni
ii) if $a\not= 0,\, b=1$, then the additional \eq\ is
satisfied if
\[ \De\psi=F_3(r,t) \q , \q \phi-\De\phi=F_4(r,t) \]

\smallskip\ni
iii) if $a\not= 0,\, b=1\pm a$, then the additional \eq\
is satisfied if
\[ \psi\pm\phi=F_5(r,t) \ . \]

We give just a simple
example for case ii), with $b=2$: observing that e.g.
$\psi=3x^2+y^2 ,\ \phi= \exp(-x)$ satisfy both \eu\ and (\ref{ci}), we
deduce
that  also
\[ \Psi=3\Big(x \cos(\lambda t)+y \sin(\lambda t)\Big)^2+
\Big(-x\sin(\la t)+y\cos(\la t)\Big)^2+\la{x^2+y^2\ov{2}} \ ,\]
\[ \Phi=\exp\Big(-(x \cos(\lambda t)+y \sin(\lambda
t))\Big)+\lambda(x^2+y^2)\]
is a family of \so s of \eu , $\forall\lambda\in R$.

It can be also shown that for no choice of $a,b$ (apart from the cases
$a=b=0$ and $a=0,b=1$ corresponding to exact \sys ) there are invariant
\so s under $X_{ab}$, then the \vf\ $X_{ab}$ is {\it not} a conditional
\sy\ for \eu .

\subsection{Example 2}

To show the usefulness of the notion of partial \sy , let us consider the
\vf
\beq X\= \psi{\pd\ov{\pd \psi}}\lb{pp} \eeq
First of all, notice that this is  trivially a
conditional \sy\ for \eu , indeed the invariance condition $X_Qu=0$ is now
$\psi=0$  and therefore simply amounts to look for the special \so s to
\eu\
with $\psi=0$. It is certainly more interesting  to show that the \vf\
(\ref{pp}) is a nontrivial partial \sy : indeed, considering this \vf\
corresponds to look for \so s to \eu\ such that the component $\psi$
admits  a
scaling property, i.e. for \so s $\psi,\phi$ such that also
$\la\psi,\phi$ solve \eu\ for all $\la\in\R$. Applying the prolongation
$X^*$ to the system \eu , and combining the resulting \eq\
$\De^{(1)}=X^*(\De)$ with \eu , one gets the new condition
\beq
[\psi,\De\psi]=0 \lb{pdp}\eeq
which is then the condition characterizing the subset ${{\cal S}}^{(1)}$
of \so s with the
above specified property. It is easy to verify that the system of the three
\eq s \eu\ and (\ref{pdp}) is symmetric under (\ref{pp}), showing that
(\ref{pp}) is indeed a partial \sy\ for \eu . Using (\ref{pdp}) one can
also rewrite the system \eu ,(\ref{pdp}) in the more convenient form
\[
{\pd\over{\pd t}}(\psi-\De\psi)+[\phi,\psi-\De\psi]+[\psi,\De\phi]\ =0 \]
\beq  {\pd\ov{\pd t}}(\De\phi)+[\phi,\De\phi]\= 0  \lb{psys}\eeq
\[ [\psi,\De\psi]=0 \]
The interesting feature of this system is that the (second) \eq\ in
(\ref{psys}) for $\phi$ is independent of $\psi$ and is equivalent to the two-dimensional 
Euler equation for an incompressible fluid, and
that, once $\phi$ is given, the  first
\eq\ for $\psi$  is  linear: obviously, this agrees with the
presence of the (partial) \sy\ given by (\ref{pp}).

Let us remark that condition (\ref{pdp}) implies that
$\De\psi=A(\psi,t)$  for some smooth function $A$, therefore the first
\eq\ in
(\ref{psys}) can be rewritten in the equivalent form ($A_\psi=\pd
A/\pd \psi$, etc.):
\[ (1-A_\psi)\Big(\psi_t-[\psi,\phi]\Big)\= A_t-[\psi,\De\phi] \ .\]
Before considering some particular cases, let us recall that our
initial system of differential \eq s \eu\ would actually contain some
physical parameters  that we have normalized to the unity up to now. In
some physical situations, however, it  can happen that one of these
parameters is negligible and then  can be put equal to zero with a
good approximation \cite{P2}: this  coefficient multiplies the term
$[\psi,\De\phi]$   in \eu\ (and in the  first \eq\ in (\ref{psys})). 

Therefore, it can be interesting to point out the following result.
{\proposition $\!\!\!$.
The truncated system
\beq \lb{a1}
{\pd\over{\pd t}}(\psi-\De\psi)+[\phi,\psi-\De\psi] \ =0 \q , \q
   {\pd\ov{\pd t}}(\De\phi)+[\phi,\De\phi]\= 0   \eeq
admits precisely the same exact \sys\ as the original one
\eu . The same is also true if one or both of the other \eq s
\beq \lb{a2}  [\psi,\De\phi]=0\q , \q [\psi,\De\psi]=0 \eeq
 are appended to the above system
(therefore, even if the partial \sy\ (\ref{pp}) is taken in
consideration also within this approximation).}

\medskip
Observing that
$[\psi,\De\phi]=0$ implies that
$\De\phi=B(\psi,t)$, the   system (\ref{a1})-(\ref{a2}) becomes
\[ (1-A_\psi)\Big(\psi_t-[\psi,\phi]\Big)\= A_t \q ,\q
\De\psi=A(\psi,t) \q , \]
\[ -B_\psi\Big(\psi_t-[\psi,\phi]\Big)=B_t \q , \q \De\phi=B(\psi,t) \ . \]
Therefore, assuming e.g. $A_\psi=1$ forces $A_t=0$; instead if
$A_\psi\not=1,B_\psi\not=0$, but $A_t=B_t=0$, the whole system (\ref{psys})
takes the very simple form
\beq \psi_t=[\psi,\phi]  \q ,\q \De\psi=A(\psi) \q , \q \De\phi=B(\psi)
\lb{AB}\eeq
Elementary \so s of  this system (and also of \eu\ and (\ref{psys}), of course) can be
immediately found by simple inspection, for instance
\[ \psi=c_1\sin(k(x-t))+c_2\sin(k(y-t))+c_3) \q , \q \phi=x-y \]
where $c_i,k$ are arbitrary constants, or also
\[ \psi=2t -\th \q , \q \phi=r^2 \]
(where as usual $\th=\arctan (y/x)$ and $r^2=x^2+y^2$), and
\[ \psi=\Psi(y-t) \q ,\q \phi=x \]
where $\Psi$ is an arbitrary regular function, just to give some
simple examples.

\section{Equations in divergence form.}

It can be interesting to remark
that our system \eu\ can be cast in the form of   divergence \eq s. For
instance, we can write \eu\ in the form (different but equivalent forms
could be also introduced)
\def\pt{\De\psi-\psi}
\beq \lb{Div} {\pd\ov{\pd t}}(\pt)+{\pd\ov{\pd x}}
\Big(\psi_y\De\phi-\phi_y\De\psi+\psi\phi_y\Big)+{\pd\ov{\pd
y}}\Big(\phi_x\De\psi-\psi_x\De\phi-\psi\phi_x\Big)\= 0\eeq
\[ {\pd\ov{\pd
t}}(\De\phi)+{\pd\ov{\pd x}}\Big(\psi_y\De\psi-\phi_y\De\phi\Big)+
{\pd\ov{\pd y}}\Big(\phi_x\De\phi-\psi_x\De\psi\Big)\= 0\]
(clearly, $\psi_x=\pd\psi/\pd x$, etc.). This property and some of its consequences 
can be summarized as follows.
{\proposition $\!\!\!$.
The system \eu\ is itself a system of conserved currents of the form
\[ {\pd J_0\ov{\pd t}}+{\rm div} \vec{J}\= 0 \ , \
{\pd K_0\ov{\pd t}}+{\rm div} \vec{K} \= 0 \]
(with clear notations from (\ref{Div})).
If  $\psi,\phi$ (together with $\De\psi,\De\phi$)  vanish  rapidly enough for
$|x|,|y|\to\infty$ 
one deduces  conservation rules for the quantities $J_0$ and $K_0$
\[ {\d \ov{\d t}}\int\!\!\int_{\R^2}(\pt)\, \d
x\d y\=  {\d \ov{\d t}}\int\!\!\int_{\R^2}\De\phi\, \d x\d y \ = 0
\lb{con}\]
and for any smooth functions thereof 
(the same is clearly true for any domain of the plane $x,y$
such that the flux of the vectors $\vec{J},\vec{K}$ through its boundary
is zero). The system (\ref{Div}), however, do not admit potential \sys .}

\medskip
The existence of these conservation rules has relevant consequences in plasma physics
(see e.g. \cite{P2}). On the other hand, 
the form (\ref{Div}) of our system suggests the possibility that \eu\ may admit {\it potential} \sys .
Indeed, according to an idea and a procedure proposed in \cite{BK},  the initial 
system \eu\ can be also written, thanks to
the form (\ref{Div}),  in a ``potential'' form, introducing a 4-dimensional
``vector potential'' $P_\a=P_\a(x,y,t)$, namely
\[ {\pd P_1\ov{\pd t}}\=\psi_x\De\phi-\phi_x(\pt)\]
\[ {\pd P_2\ov{\pd t}}\=\psi_y\De\phi-\phi_y(\pt)\]
\[ -{\pd P_2\ov{\pd x}}+{\pd P_1\ov{\pd y}}\=\pt \]
\[ {\pd P_3\ov{\pd t}}\=\psi_x\De\psi-\phi_x\De\phi\]
\[ {\pd P_4\ov{\pd t}}\=\psi_y\De\psi-\phi_y\De\phi\]
\[ -{\pd P_4\ov{\pd x}}+{\pd P_3\ov{\pd y}}\=\De\phi \]
where the ``unknown'' functions are the six
quantities $\psi,\phi$ and $P_\a$.
In principle, one could look for Lie \sys\ for this system, but either
direct
calculations or an extension of a result given in \cite{PS}, which
fixes precise conditions
on the order of derivatives appearing in the \eq s, can show that it
does not admit \sys\ exhibiting explicit dependence on the vector
potential $P_\a$;
this implies (see \cite{BK,PS}) that there are no potential
(nonlocal) \sys\ for our
problem  in its original form \eu .

\section{Conclusions.}

As mentioned in the Introduction, the set of equations (\ref{EU}) are 
of the interest  for the
study of collisionless magnetic field line reconnection in a plasma. 
In particular, some of
the explicit solutions found in the present article using  Lie point 
symmetries have a direct
physical interpretation,  e.g. in the case of the solutions 
(\ref{VW})  as plasma
configurations forced from the boundaries. The  investigation  of the 
physical properties of
these configurations will be the subject of a forthcoming article.

\vfill\eject

\end{document}